\documentclass{article}
\usepackage{spconf,amsmath,graphicx}
\usepackage{enumitem}
\usepackage{subfig}
\usepackage{graphicx}
\usepackage{caption}
\usepackage{scrextend}
\usepackage{color,soul}
\usepackage{cite}
\usepackage[colorlinks=true]{hyperref}
\usepackage[symbol]{footmisc}
\include{IEEEtran}

\title{BBAND Index: A No-reference Banding Artifact Predictor}
\name{Zhengzhong~Tu$^{\star}$\thanks{$^{\star}$Work done during an internship at Google.},~Jessie~Lin$^{\dagger}$,~Yilin~Wang$^{\dagger}$,~Balu~Adsumilli$^{\dagger}$,~and~Alan~C.~Bovik$^{\star}$}
\address{$^{\star}$Laboratory for Image and Video Engineering (LIVE), The University of Texas at Austin \\
$^{\dagger}$YouTube Media Algorithms Team, Google Inc.}

\begin{document}
%
\maketitle
\begin{abstract}
Banding artifact, or false contouring, is a common video compression impairment that tends to appear on large flat regions in encoded videos. These staircase-shaped color bands can be very noticeable in high-definition videos. Here we study this artifact, and propose a new distortion-specific no-reference video quality model for predicting banding artifacts, called the \textbf{B}lind \textbf{BAN}ding \textbf{D}etector (BBAND index). BBAND is inspired by human visual models. The proposed detector can generate a pixel-wise banding visibility map and output a banding severity score at both the frame and video levels. Experimental results show that our proposed method outperforms state-of-the-art banding detection algorithms and delivers better consistency with subjective evaluations. 
\end{abstract}

\section{INTRODUCTION}
\label{sec:intro}
Banding/false contour remains one of the dominant artifacts that plague the quality of high-definition (HD) videos, especially when viewed on high-resolution or Retina displays. Yet, while significant research effort has been devoted to analyzing various specific compression related artifacts \cite{shahid2014no}, such as noise \cite{norkin2018film}, blockiness \cite{wang2000blind}, ringing \cite{marziliano2004perceptual}, and blur \cite{marziliano2002no}, less attention has been paid to analyzing banding/false contours. Given the rapidly growing demand for HD/Ultra-HD videos, the need to assess and mitigate banding artifacts is receiving increased attention in both academia and industry.

\begin{figure}[!t]
\subfloat[Original UGC]{
    \includegraphics[width=0.225\textwidth]{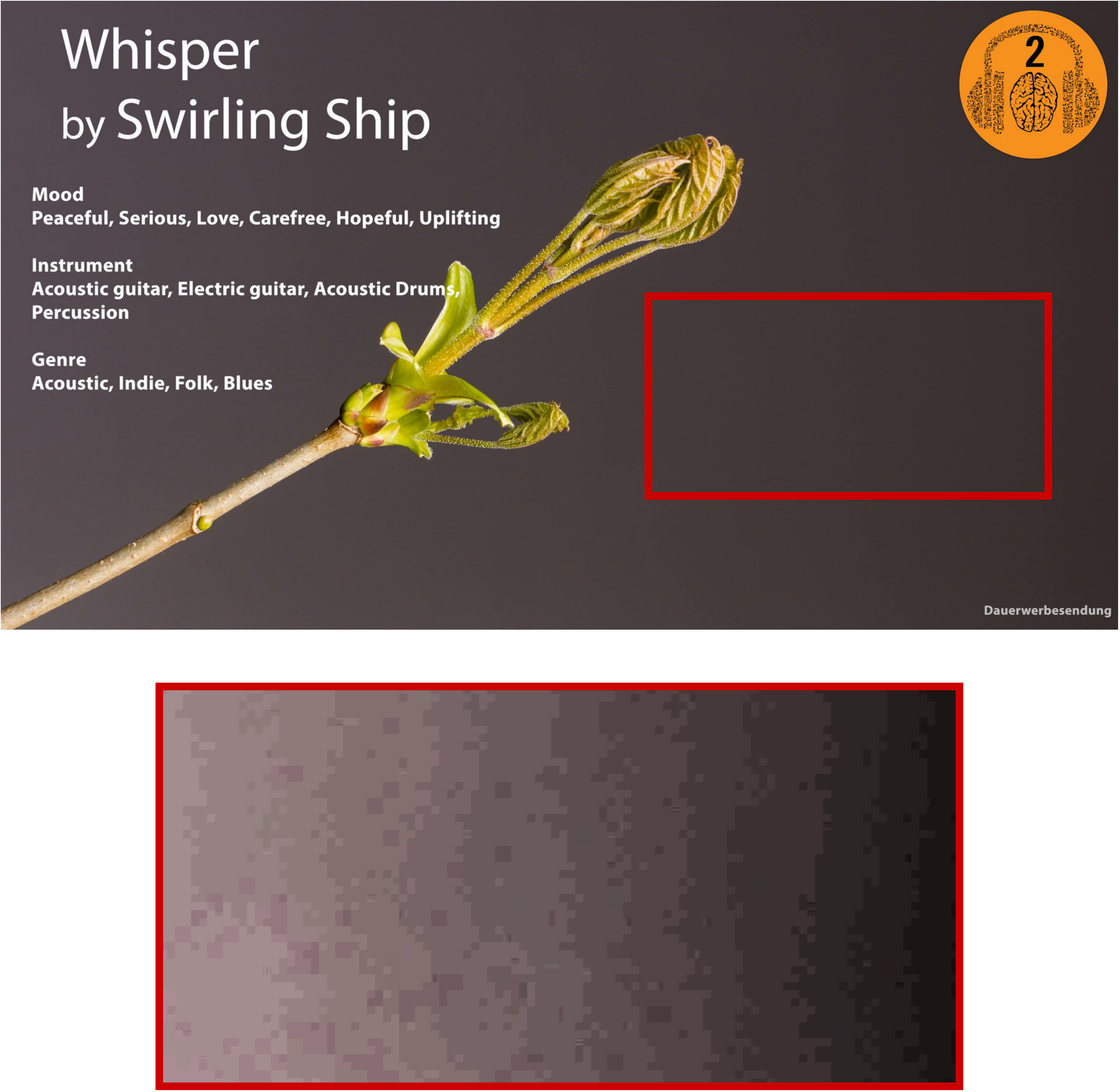}
    \label{orig}}
    \hfill
\subfloat[Transcoded/Re-encoded]{
    \includegraphics[width=0.225\textwidth]{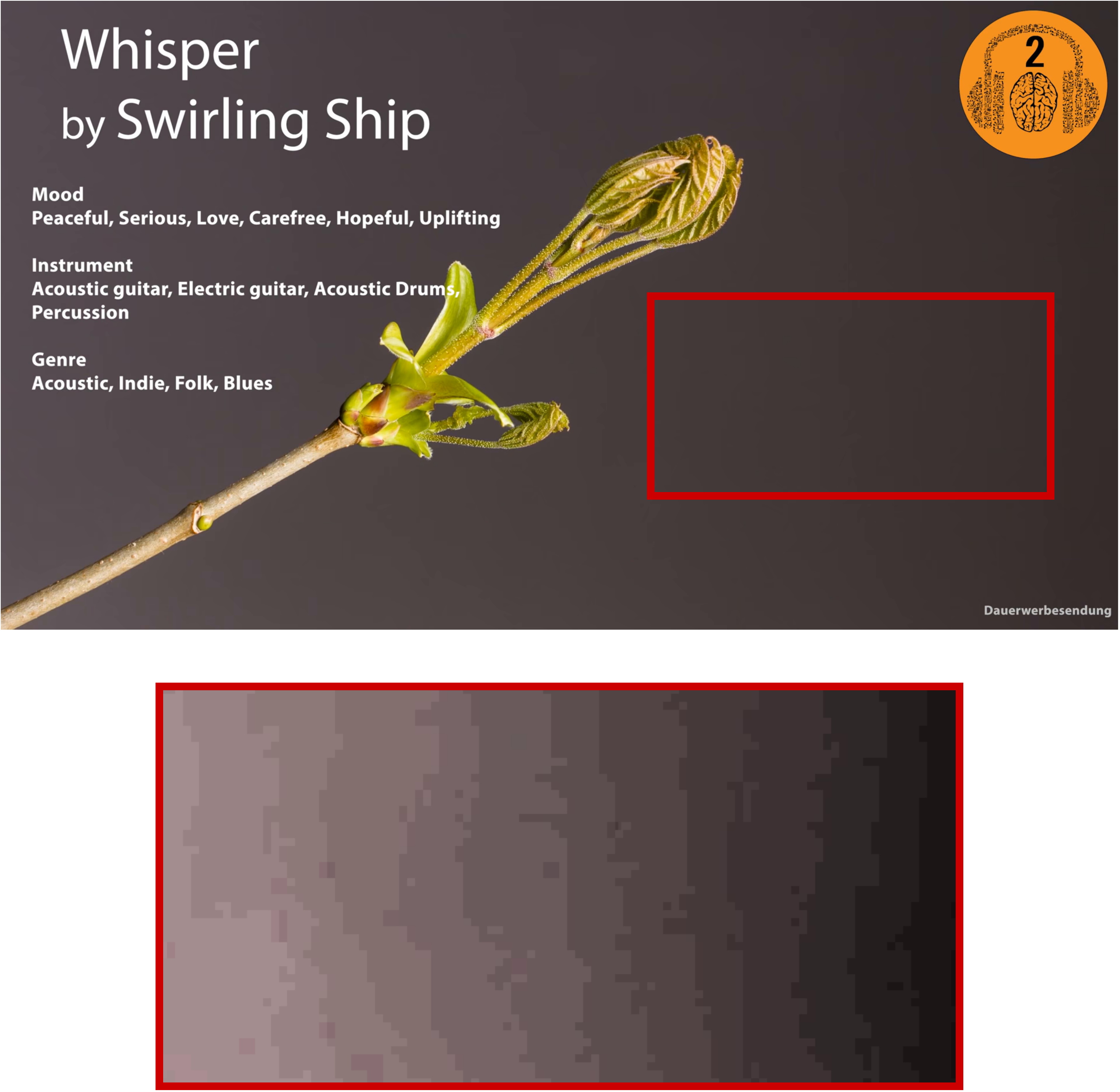}
    \label{tran}}
    \hfill
\caption{Banding artifacts exacerbated by transcoding/re-encoding. (a) shows a frame sampled from an original UGC video with less noticeable ``noisy'' banding edges, while VP9-encoding exhibits more visible ``clean'' banding edges, as shown in (b). The lower figures show contrast-enhanced banding regions for better visualization.}
\label{fig:bandeg}
\end{figure}

\begin{figure*}[!ht]
\centering
\includegraphics[width=0.89\textwidth]{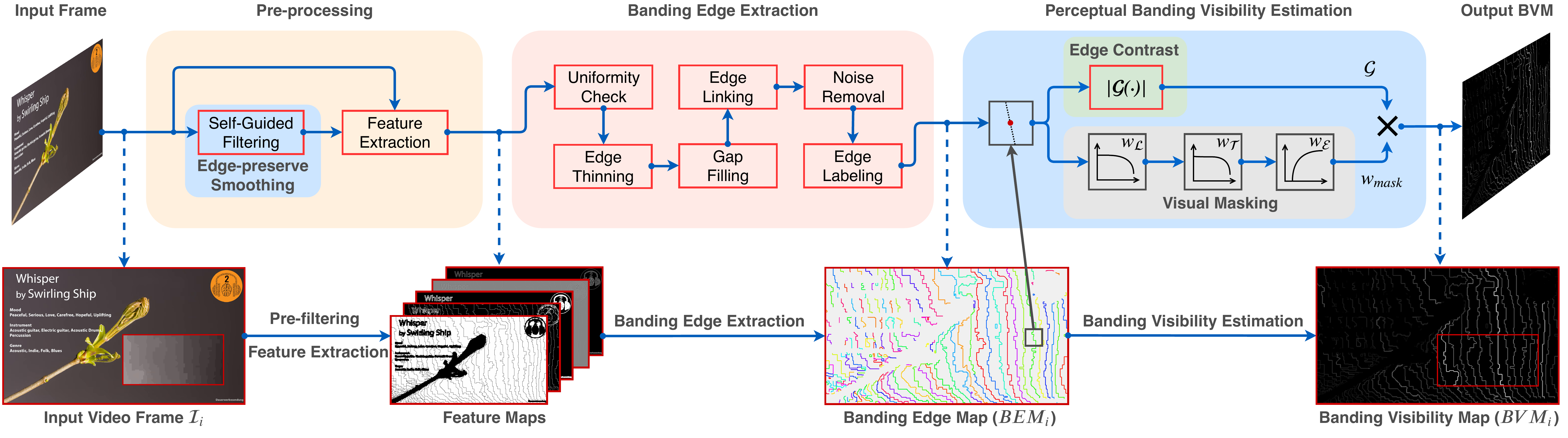}
\caption{Schematic overview of the first portion (Section \ref{ssec:pre-proc}-\ref{ssec:perc_vis_est}) of the proposed BBAND model. The first row shows the processing flow, while the second row depicts exemplar responses of each processing block.}
\label{fig:bbad_flowchart}
\end{figure*}

Banding appears in large, smooth regions with small gradients and presents as discrete, often staircased bands of brightness or color as a result of quantization in video compression. All popular video encoders, including H.264/AVC \cite{wiegand2003overview}, VP9 \cite{mukherjee2013latest}, and H.265/HEVC \cite{sullivan2012overview} can introduce these artifacts at lower or medium bitrate when coding contents containing smooth areas. Fig. \ref{fig:bandeg} shows an example of banding artifacts exacerbated by transcoding. Traditional quality prediction algorithms such as PSNR, SSIM \cite{wang2004image}, and VMAF \cite{li2016toward}, however, do not align well with human perception of banding \cite{wang2016perceptual}. The development of a highly reliable banding detector for both original user-generated content (UGC) and transcoded/re-encoded videos, would, therefore, greatly assist streaming platforms in developing measures to avoid banding artifacts in streaming videos.

\textbf{Related Work.} There exists some prior study relating to banding/false contour detection. Some methods \cite{daly2004decontouring, lee2006two,  huang2016understanding} take advantage of local features such as the gradient, contrast or entropy to measure potential banding edge statistics. However, methods like these generally do not perform very well when applied to assess the severity of banding edges in video content. Another approach to banding detection is based on pixel segmentation \cite{bhagavathy2009multiscale, baugh2014advanced, wang2016perceptual}, where a bottom-up connected component analysis is used to first detect uniform segments, usually followed by a process of banding edge separation. These methods are often sensitive to edge noise, though. We do not include block-based processing, as in \cite{jin2011composite, wang2014multi}, since it is hard to classify blocks where banding and textures coexist. If post-filtering is applied to these blocks, textures near the banding may become over-smoothed.

Our objective is to design an adaptive blind processor which can detect or enhance both ``noisy'' banding artifacts that arise in original UGC videos, as well as ``clean'' banding edges in transcoded videos. In this regard, it could be utilized as a basis for the development of pre-processing and post-processing \textit{debanding} algorithms. More recent banding detectors like the False Contour Detection and Removal (FCDR) \cite{huang2016understanding} and Wang's method \cite{wang2016perceptual} are not designed for this practical purpose, and hence it is essential to devote more research to developing other adaptive banding predictors applicable to pre- or post-debanding implementations.

%
In this paper we propose a new, ``completely blind'' \cite{mittal2012making} banding model, dubbed the \textbf{B}lind \textbf{BAN}ding Artifact \textbf{D}etector (BBAND index), by leveraging edge detection and a human visual model. The proposed method operates on individual frames to obtain a pixel-wise banding visibility map. It can also produce no-reference perceptual quality predictions of videos with banding artifacts. Details of our proposed banding detector are given in Section \ref{sec:proposed}, while evaluation results are given in Section \ref{sec:subj_eval}. Finally, Section \ref{sec:conc} concludes the paper.

\section{Proposed Banding Detector}
\label{sec:proposed}
A block diagram of the first portion of the proposed model, which generates a pixel-wise banding visibility map (BVM), is illustrated in Fig. \ref{fig:bbad_flowchart}. Based on our observation that banding artifact appears as weak edges with small gradient (whether ``clean'' or ``noisy''), we build our banding detector (BBAND), by exploiting existing edge detection techniques as well as certain visual properties. A spatio-temporal visual importance pooling is then applied to the BVM, as shown in Fig. \ref{fig:pool_flowchart},  yielding ``completely blind'' banding scores for both individual frames and the entire video.
\renewcommand{\thefootnote}{\fnsymbol{footnote}}
\subsection{Pre-processing}
\label{ssec:pre-proc}
We have observed that re-encoding videos at bitrates optimized for streaming often exacerbates banding in videos that already exhibit slight banding artifacts that may be barely visible, as shown in Fig. \ref{fig:bandeg}\footnote[4]{The exemplary frames used in this paper are from Music2Brain (YouTube channel). Website: https://bigweb24.de. Used with permission.\label{fn1}}. We thereby deployed self-guided filtering \cite{he2012guided},  which is an effective edge-preserving smoothing process, to enhance banding edges. We deemed the guided to be a better choice than the bilateral filter \cite{tomasi1998bilateral}, since it better preserves gradient profile, which is a vital local feature used in our proposed framework. Image gradients are then calculated by applying a Sobel operator after pre-smoothing, yielding a gradient feature map. 

\subsection{Banding Edge Extraction}
\label{ssec:band_edge_extract}
Inspired by the efficacy of using the Canny edge detector \cite{canny1986computational} to improve ringing region detection \cite{liu2010perceptually}, we performed a similar procedure to extract banding edges. After pre-filtering, the pixels are classified into three classes depending on their Sobel gradient profiles: pixels having Sobel gradient magnitudes less than $T_1$ are labeled as flat pixels; pixels with gradient magnitudes exceeding $T_2$ are marked as textures. The remaining pixels are regarded as candidate banding pixels (CBP), on which the following steps are implemented to create a banding edge map (BEM). (We used $(T_1,T_2)=(2,12)$).
\noindent
\begin{enumerate} [label= \arabic{enumi}.,ref=Step \arabic{enumi}, leftmargin=*,topsep=0pt,itemsep=-1ex,partopsep=1ex,parsep=1ex]
    \item \textit{Uniformity Check}: Only the CBPs whose neighbors are either flat pixels or CBPs are retained for further processing.
    \item \textit{Edge Thinning}: Non-maxima suppression \cite{canny1986computational} is applied to each remaining CBP along its Sobel gradient orientation to better localize the potential bands.
    \item \textit{Gap Filling}: If two candidate pixels are disjoint, but able to be overlapped by a binary circular blob, the gap between the two points is filled by a proper banding edge.
    \item \textit{Edge Linking}: All connected CBPs are linked together in lists of sequential edge points. Each edge is either a curved line or a loop.
    \item \textit{Noise Removal}: Linked edges shorter than a certain threshold are discarded as visually insignificant.
    \item \textit{Edge Labeling}: The resulting connected banding edges are labeled separately, defining the ultimate BEM.
\end{enumerate}

The colored edge map in Fig. \ref{fig:bbad_flowchart} shows a BEM extracted from an input frame. The banding edges are well localized.

\subsection{Banding Visibility Estimation}
\label{ssec:perc_vis_est}
Staircase-like banding artifacts appear similar to Mach Bands (Chevreul illusion), where perceived edge contrast is exaggerated by edge enhancement by the early visual system \cite{wiki:machbands}. Explanations of the illusion usually involve the center-surround excitatory-inhibitory pooling responses of retinal receptive fields  \cite{ratliff1965mach}. Inspired by the psychovisual findings in \cite{ross1989conditions}, we developed a local banding visibility estimator based on edge contrast and perceptual masking effects. The estimator processes the BEM and yields an element-wise banding visibility map (BVM).
\subsubsection{Basic Edge Feature}
\label{sssec:edge_feat}
Banding artifact presents as visible edges. As described earlier, we use the Sobel gradient magnitude as an edge visibility feature. Since edge visibility is also affected by content, we also model visual masking as it may affect the subjective perception of banding.
\subsubsection{Visual Masking}
\label{sssec:visual_mask}
Visual masking is a phenomenon whereby the visibility of a visual stimulus (target) is reduced by the presence of another stimulus, called a mask. Well-known masking effects include luminance and texture masking \cite{liu2010perceptually, chen2016perceptual}. Here we deploy a simple but effective quantitative model of the effect of masking on banding edge visibility.

\begin{figure}[!t]
\centering
\includegraphics[width=0.48\textwidth]{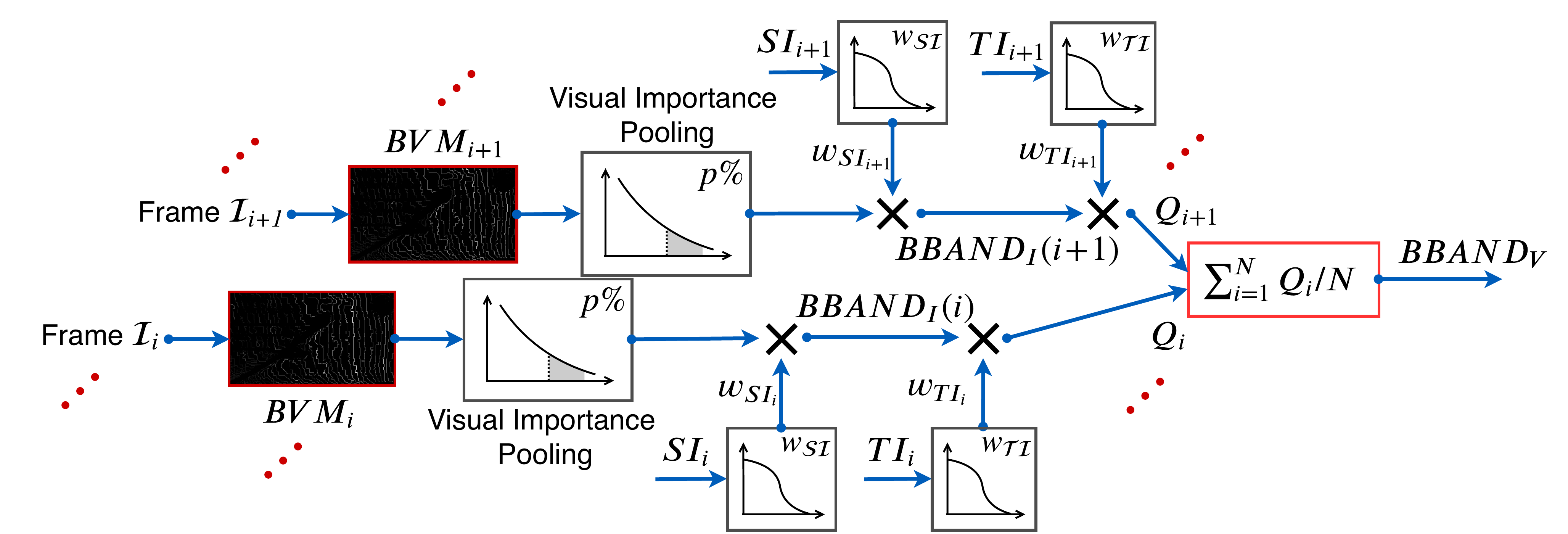}
\caption{Flowchart of the second portion (Section \ref{ssec:band_metric}) of the proposed BBAND model, which produces banding scores on both frames and whole videos.}
\label{fig:pool_flowchart}
\end{figure}

\noindent\textbf{Local Statistics}: At each detected banding pixel in the BEM, compute local Gaussian-weighted mean and standard deviation (``sigma field'') on the original un-preprocessed frame:
\begin{equation}
\label{eq:mu}   
\mu(i,j)=\sum_{k=-K}^K\sum_{\ell=-L}^Lw_{k,\ell}\mathcal{I}(i-k,j-\ell)
\end{equation}
\begin{equation}
\label{eq:sigma}
\sigma(i,j)=\sqrt{\sum_{k=-K}^K\!\sum_{\ell=-L}^Lw_{k,\ell}[\mathcal{I}(i\!-\!k,j\!-\!\ell)-\mu(i,j)]^2},
\end{equation}
where $(i,j)$ are spatial indices at detected pixels in the BEM with corresponding original pixel intensity $\mathcal{I}(i,j)$, and $\boldsymbol{w}=\{w_{k,\ell}|k=-K,...,K,\ell=-L,...,L\}$ is a 2D isotropic Gaussian weighting function. We use the $\mu(\cdot)$ and $\sigma(\cdot)$ feature maps to estimate the local background luminance and complexity. The window size in our experiments was set as $9\!\times\!9$. 


\noindent\textbf{Luminance Masking}: We define a luminance visibility transfer function ($\mathrm{VTF}_\ell$) to express luminance masking as a function of the local background intensity. We have observed that banding artifacts remain visible even in very dark areas, so we only model the masking at very bright pixels. A final luminance masking weight is computed at each pixel as
\begin{equation}
\label{eq:vtf_l}
w_{\ell}(i,j)=\left\{
\begin{array}{lll}
    1 & \mu(i,j)\leq \mu_0 \\
    1-\alpha(\mu(i,j)-\mu_0)^\beta & \mu(i,j)>\mu_0,
\end{array}
\right.
\end{equation}
where $\mu(i,j)$ is calculated using (\ref{eq:mu}). $(\alpha,\beta)$ is a pair of constants chosen to adjust the shape of the transfer function. We used $(\alpha,\beta,\mu_0)=(1.6\!\times\!10^{-5},2,81)$ in our implementations.

\noindent\textbf{Texture Masking}: We also define a texture visibility transfer function ($\mathrm{VTF}_t$) to capture the effects of texture masking. The $\mathrm{VTF}_t$ is defined to be inversely proportional to local image activity \cite{liu2010perceptually} when an activity measure (mean ``sigma field'') rises above threshold $\lambda_0$. The overall weighting function is formulated as
\begin{equation}
\label{eq:vtf_t}
w_{t}(i,j)=\left\{
\begin{array}{ll}
    1 & \lambda(i,j)\leq \lambda_0 \\
    1/[1+(\lambda(i,j)-\lambda_0)]^\gamma & \lambda(i,j)>\lambda_0,
\end{array}
\right.
\end{equation}
and
\begin{equation}
\label{eq:lambda}
\lambda(i,j)=\frac{1}{(2K\!+\!1)(2L\!+\!1)}\sum_{k=-K}^K\sum_{\ell=-L}^L\sigma(i-k,j-\ell),
\end{equation}
where $\sigma(i,j)$ is given by Eq. (\ref{eq:sigma}), and $\gamma$ is a parameter that is used to tune the nonlinearity of $\mathrm{VTF}_t$. The values of $(\gamma,\lambda_0)=(5,0.32)$ were adopted after careful inspection.

\noindent\textbf{Cardinality Masking}: The authors of \cite{wang2016perceptual} have shown that edge length is another useful banding visibility feature in a subjective study. We accordingly define the following transfer function which weights banding visibility by edge cardinality:
\begin{equation}
\label{eq:vtf_c}
w_{c}(i,j)=\left\{
\begin{array}{ll}
    0 & |\mathrm{E}(i,j)|\leq c_0 \\
    (|\mathrm{E}(i,j)|/\sqrt{MN})^\eta & |\mathrm{E}(i,j)|> c_0,
\end{array}
\right.
\end{equation}
where $\mathrm{E}(i,j)=\{\mathrm{E}\in \mathrm{BEM}|(i,j)\in \mathrm{E}\}$ is the set of banding edges passing through location $(i,j)$, and $c_0$ is a threshold on minimal noticeable edge length, above which banding edge visibility is positively correlated to normalized edge length. $M$ and $N$ denote the image height and width, respectively. We used parameters $(c_0,\eta)=(16,0.5)$ in our experiments.
\begin{figure*}[!ht]
\centering
\subfloat[Baugh \cite{baugh2014advanced}]{
    \includegraphics[width=0.24\textwidth]{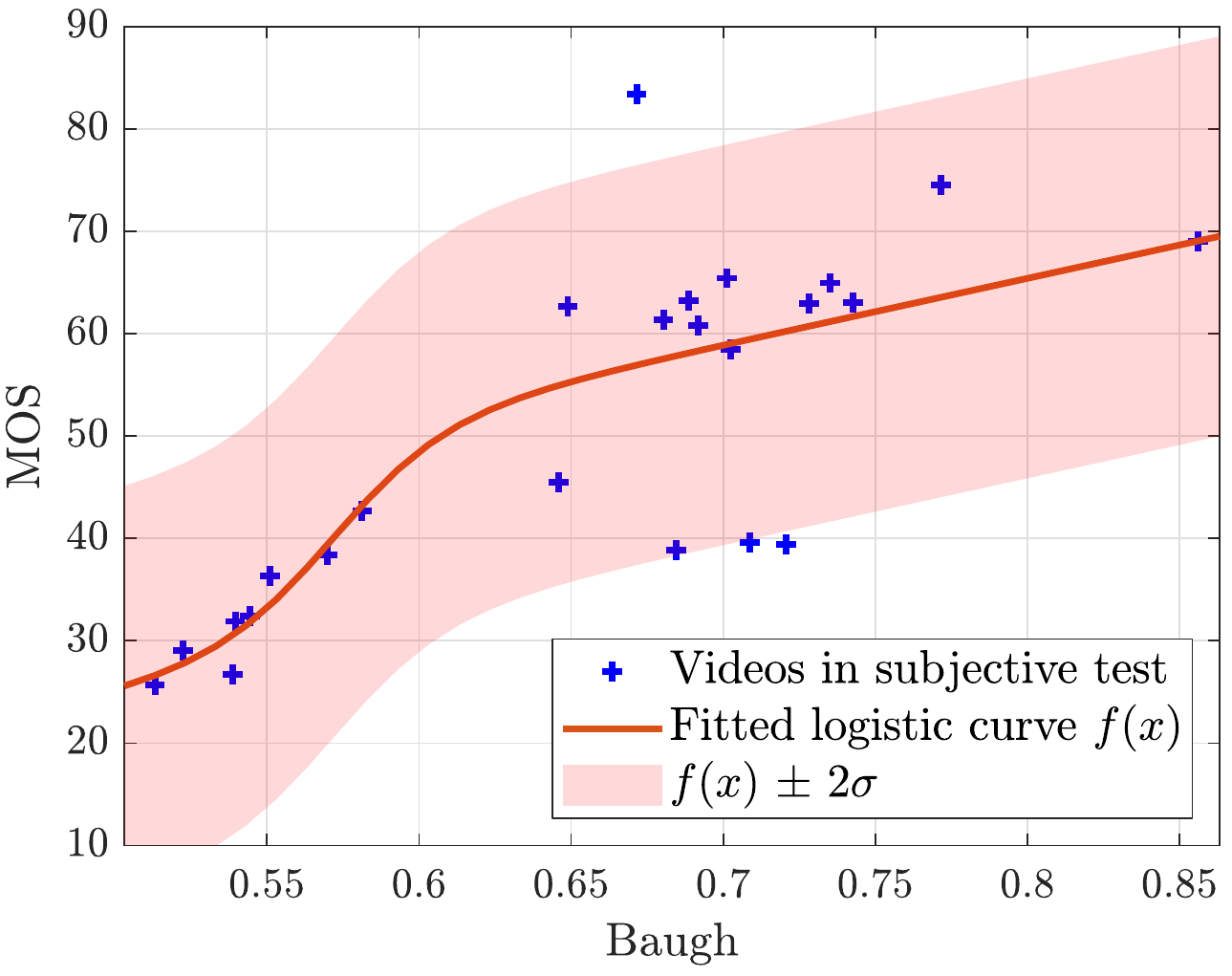}
    \label{baugh}}
    \hfil
\subfloat[Wang \cite{wang2016perceptual}]{
    \includegraphics[width=0.24\textwidth]{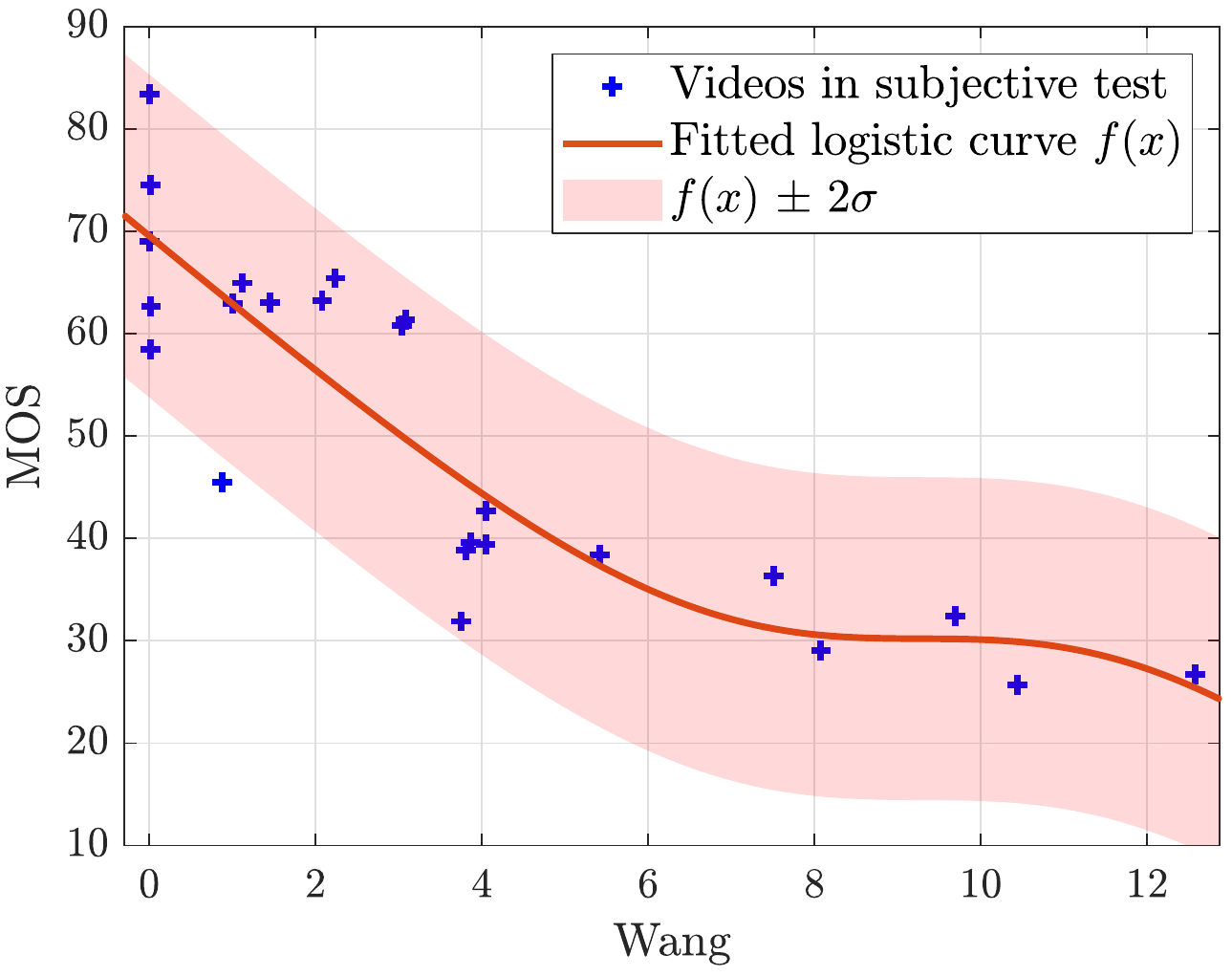}
    \label{wang}}
    \hfil
\subfloat[BBAND (proposed)]{
    \includegraphics[width=0.24\textwidth]{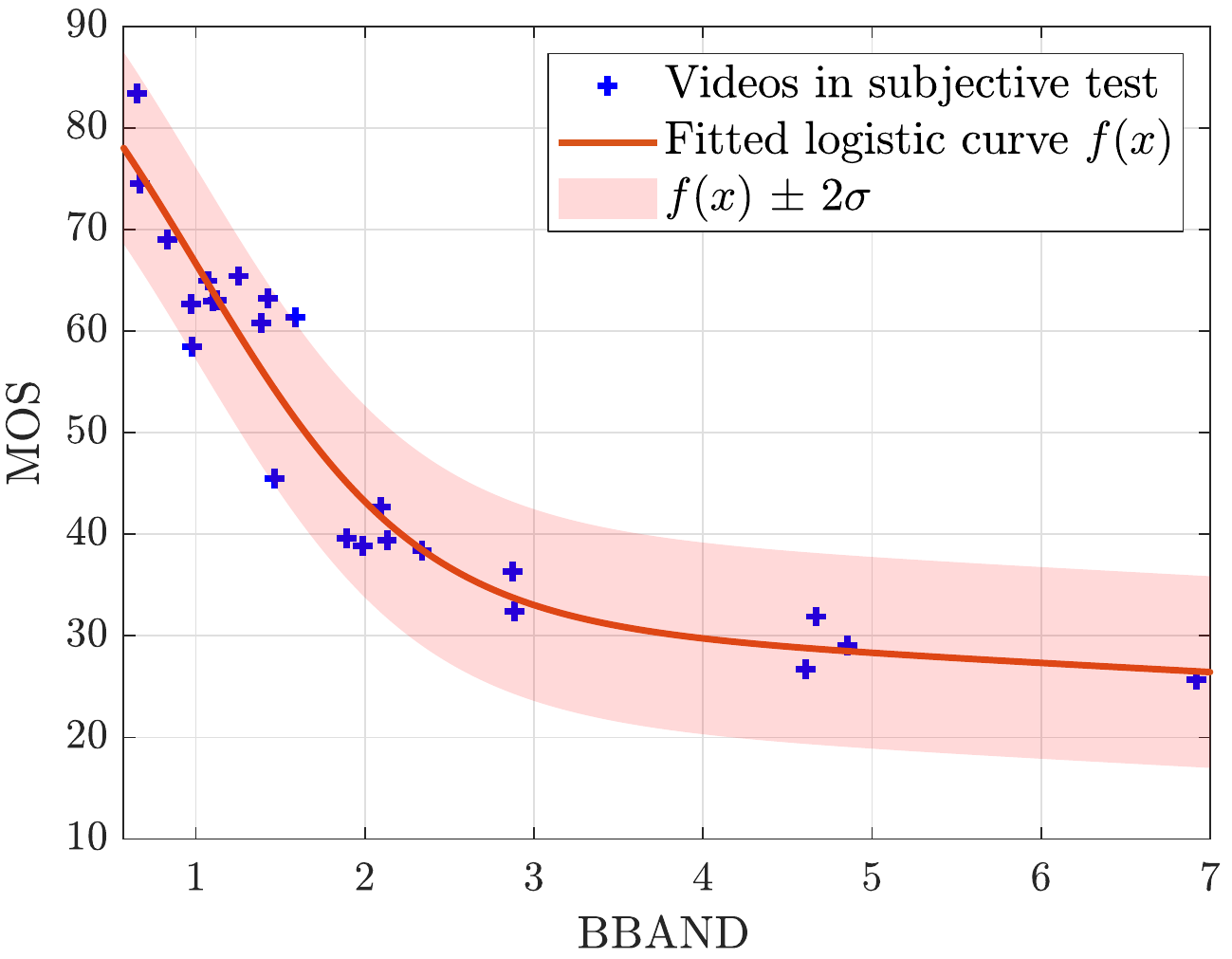}
    \label{bbad}}
\caption{Scatter plots and regression curves of (a) Baugh \cite{baugh2014advanced}, (b) Wang \cite{wang2016perceptual}, (c) BBAND, versus MOS on banding dataset \cite{wang2016perceptual}.}
\label{fig3:scat_plot}
\end{figure*}
\subsubsection{Visibility Integration}
\label{sssec:int_strategy}
The overall visibility of an artifact depends on the visual response to it modulated by a concurrency of masking effects. Here we use a simple but effective product model of feature integration at each computed banding pixel to obtain the banding visibility map:
\begin{equation}
\label{eq:bvm}
\mathrm{BVM}(i,j)=w_{\ell}(i,j)\cdot w_{t}(i,j)\cdot w_{c}(i,j)\cdot |\mathcal{G}(i,j)|,
\end{equation}
where $w(\cdot)$'s are the responsive weighting parameters that scale the measured edge strength (Sobel gradient magnitude) $|\mathcal{G}(i,j)|$ at location $(i,j)$.

\subsection{Making a Banding Metric}
\label{ssec:band_metric}
Previous authors \cite{chen2016perceptual, ghadiyaram2017no, moorthy2009visual, park2012video} have studied the benefits of integrating visual importance pooling into objective quality model, generally aligning with the idea that the overall perceived quality of a video is dominated by those regions having the poorest quality. In our model, we apply the worst $p\%$ percentile pooling to obtain an average banding score from the extracted BVM, where $p=80$ is employed in experiments.

\begin{table}
\renewcommand{\arraystretch}{1}
\caption{Performance comparison of blind banding models.}
\label{table:corr}
\centering
\begin{tabular}{lcccc}
\hline
Metric &  SRCC &  KRCC &  PLCC & RMSE \\
\hline\hline
Baugh \cite{baugh2014advanced} & 0.7739 & 0.6304 & 0.8037 & 9.7671\\
Wang \cite{wang2016perceptual} & 0.8689 & 0.6788 & 0.8770 & 7.8863\\
BBAND & \bfseries 0.9330 & \bfseries 0.8116 & \bfseries 0.9578 & \bfseries 4.7173\\
\hline
\end{tabular}
\end{table}

Banding usually occurs in non-salient regions (e.g., background) while salient objects catch more of the viewer's attention. We thereby use the well-known spatial information (SI) and temporal information (TI) to indicate possible spatial and temporal distractors against banding visibility. SI is computed as the standard deviation of the pixel-wise gradient magnitude, while TI as the standard deviation of the absolute frame differences on each frame \cite{itu1999subjective}. These are then mapped by an exponential transfer function to obtain weights:
\begin{equation}
\label{eq:si}
w_{i}(x)=\exp{(-a_ix^{b_i})},\ i\in\{\mathrm{SI}, \mathrm{TI}\}.
\end{equation}

Finally, we construct the frame-level BBAND index by applying visual percentile pooling and $\mathrm{SI}$ weights to BVM:
\begin{equation}
\label{eq:bbad_i}
\resizebox{.9\hsize}{!}{$\mathrm{Q}_\mathrm{BBAND_\mathcal{I}}(\mathcal{I})=w_{\mathrm{SI}}(\mathrm{SI})\!\cdot\!\cfrac{1}{|\mathcal{K}_{p\%}|}{\displaystyle\sum}_{(i,j)\in \mathcal{K}_{p\%}}\!\mathrm{BVM_\mathcal{I}}(i,j)$},
\end{equation}
where $\mathcal{K}_{p\%}$ is the index set of the largest $p^{th}$ percentile non-zero pixel-wise visibility values contained in the BVM of frame $\mathcal{I}$. We also obtain the video-level BBAND metric by averaging all frame-level banding scores, weighted by per-frame TI, respectively:
\begin{equation}
\label{eq:bbad_v}
\resizebox{.86\hsize}{!}{$\mathrm{Q}_\mathrm{BBAND_\mathcal{V}}(\boldsymbol{\mathcal{V}})=\cfrac{1}{N}{\displaystyle\sum}_{n=1}^N w_{\mathrm{TI}}(\mathrm{TI}_n)\cdot \mathrm{Q}_{\mathrm{BBAND_\mathcal{I}}}(\mathcal{I}_n)$}.
\end{equation}

Fig. \ref{fig:pool_flowchart} shows the entire workflow of the BBAND indices.

\section{Subjective Evaluation}
\label{sec:subj_eval}
Other implemented parameters in our proposed BBAND model are  $(a_{\mathrm{SI}},b_{\mathrm{SI}},a_{\mathrm{TI}},b_{\mathrm{TI}})\!=\!(10^{-6},3,2.5\times10^{-3},2)$, respectively, after empirical calibration, and we've found these selected parameters generally perform well in most cases. We evaluated the BBAND model against two recent banding metrics, Wang \cite{wang2016perceptual} and Baugh \cite{baugh2014advanced}, on the only existing banding dataset, created by Wang et al. \cite{wang2016perceptual}. It consists of six clips of 720p@30fps videos with different levels of quantization using VP9. The Spearman rank-order correlation coefficient (SRCC) and Kendall rank-order correlation coefficient (KRCC) between predicted scores and mean opinion scores (MOS) of subjects are directly reported for the evaluated methods. We also calculated the Pearson linear correlation coefficient (PLCC) and the corresponding root mean squared error (RMSE) after fitting a logistic function between MOS and predicted values \cite{sheikh2006statistical}. Table \ref{table:corr} summarizes the experimental results, and Fig. \ref{fig3:scat_plot} plots the fitted logistic curves of MOS versus the evaluated banding models. These results have shown that the proposed BBAND metric yields highly promising performance regarding subjective consistency. 

\section{Conclusion and Future Work}
\label{sec:conc}
We have presented a new no-reference video quality model called the BBAND for assessing perceived banding artifacts in high-quality or high-definition videos. The algorithm involves robust detection of banding edges, a perception-inspired estimator of banding visibility, and a model of spatial-temporal visual importance pooling. Subjective evaluation shows that our proposed method correlates favorably with human perception as compared to several existing banding metrics. As a ``completely blind'' (opinion-unaware) distortion-specific quality indicator, BBAND can be incorporated with other video quality measures as a tool to optimize user-generated video processing pipelines for media streaming platforms. Future work will include further improvements of BBAND by integrating with more temporal cues, and its applications to address such banding artifacts via debanding pre-processing or post-filtering.




\vfill\pagebreak

\section{REFERENCES}
\label{sec:refs}


\bibliographystyle{IEEEtran}
\small\bibliography{IEEEabrv,refs}

\end{document}